\documentstyle[preprint,prl,aps]{revtex}
\tightenlines
\begin{document}
\draft
\title{Extended King Models for Star Clusters}
\author{Kwok Sau Fa and I. T. Pedron \\
%EndAName
Departamento de f\'{\i}sica, Universidade Estadual de Maring\'{a},\\
Av. Colombo, 5790, 87020-900 - Maring\'{a} - PR, Brazil}
\maketitle

\begin{abstract}
Models applied to galaxies in equilibrium configuration are based on the
solution of the collisionless Boltzmann equation and the Poisson equation
for gravitational interaction which are related to each other by a
smoothed-out mass density. This mass density is related to a distribution
function $f(\mathbf{x},\mathbf{v},t)$ and produces a gravitational potential 
$\phi (\mathbf{x},t)$. The most popular models to describe spherically
symmetric systems are King models. These models are based on a truncated
isothermal sphere. They fit the surface brightness of globular clusters and
some elliptical galaxies well. A well-known example of a galaxy that King
models does not fit well is NGC 3379. In this work, we extend King models.
Our models are based on the Tsallis distribution function. There is a
parameter $q$ in this function and we can recover the King
distribution function in the limit $q\rightarrow 1$. We discuss the general behaviors
of our models and, moreover, we use them to fit the surface brightness of
NGC 3379 and 47 TUC.

\medskip \medskip

PACS: 05.90.+m, 98.62.Ve, 98.56.Ew

\newpage
\end{abstract}

A lot of models for star clusters, based on the collisionless Boltzmann
equation and Poisson equation, have been studied in the last century \cite
{binney,padma}. In this approach, the state of the system is described by a
distribution function $f(\mathbf{x},\mathbf{v},t)$ in phase space and
considering the smooth gravitational potential. These models can be applied
to star clusters in equilibrium configuration or when the characteristic
time of these structures does not exceed the relaxation time \cite{binney}.

The most commonly used models to describe spherically symmetric systems are
King models. These models can fit the brightness distributions of globular
clusters and some elliptical galaxies well \cite
{binney,padma,king,kormendy,mihalas}. King models are based on the following
distribution function

\begin{equation}
f_{K}(\varepsilon )=\left\{ 
\begin{array}{c}
\rho _{1}\left( 2\pi \sigma ^{2}\right) ^{-3/2}\left( e^{\varepsilon /\sigma
^{2}}-1\right) \text{ \ \ \ \ }\varepsilon >0 \\ 
\text{ \ \ \ \ }0\text{ \ \ \ \ \ \ \ \ \ \ \ \ \ \ \ \ \ \ \ \ \ \ \ \ \ \
\ \ \ \ \ \ \ }\varepsilon \leq 0
\end{array}
\right. \text{ .}  \label{eq1}
\end{equation}
where $\varepsilon $ is the relative energy $\varepsilon =-E+\Phi _{0}=\Psi
-v^{2}/2$ and $\Psi $ is the relative potential $\Psi =-\Phi +\Phi _{0}$.
This distribution function may be viewed as a truncated isothermal sphere.
The characteristic of this model is that the relative potential $\Psi $ and
density of stars $\rho $ become equal to zero at some finite radius $r_{t}$
(tidal radius). The value of $r_{t}$ depends on the central potential $\Psi
(0)/\sigma ^{2}$ and King radius defined as $r_{0}=\sqrt{9\sigma ^{2}/4\pi
G\rho _{0}}$. In the limit $\Psi (0)/\sigma ^{2}\rightarrow \infty $, the
isothermal sphere is recovered and $r_{t}$ it extends to infinity.

In this work, we generalize King models. The main goal of this
generalization is to try to describe some elliptical galaxies where King
models cannot describe them  well. Thus, we are also restricted to
spherically symmetric models and the distribution function depends only on
relative energy $f(\varepsilon )$. These kinds of models possess many
solutions for the collisionless Boltzmann equation: Jeans stated that any
steady-state solution can be expressed through integrals of motion on
phase-space coordinates. Moreover, all these models are stable \cite{binney}.

Our distribution function is based on Tsallis' statistics \cite{tsallis},
with the entropy and distribution function given by $S=k(1-%
\sum_{i}p_{i}^{q})/(1-q)$ and $f_{T}(\varepsilon )=k\left( 1+\beta
(1-q)\varepsilon \right) ^{1/(1-q)}$, respectively. In this type of
statistics, $q$ represents a parameter which describes different statistics.
It has been successfully applied to several systems such as L\'{e}vi-type
anomalous superdiffusion \cite{levy}, Euler turbulence \cite{bogo}, and
anomalous relaxation electron-phonon interaction \cite{koponen}. In
astrophysics and cosmology we can mention the following set of works \cite
{plastino}; but for our purposes, it is worthwhile to mention the work of
Plastino and Plastino \cite{plastino}. They have shown that Tsallis' entropy
can determine the meaningful distribution function as described by
polytropic models. Therefore, our distribution function, based on Tsallis'
distribution function, is chosen as follows:

\begin{equation}
f(\varepsilon )=\left\{ 
\begin{array}{c}
\rho _{1}\left( 2\pi \sigma ^{2}\right) ^{-3/2}\left\{ \left[
1+(1-q)\frac{\varepsilon}{\sigma ^{2}} \right] ^{\frac{1}{1-q}}-1\right\} \text{ \ \ \ \ }%
\varepsilon >0 \\ 
\text{ \ \ \ \ \ \ \ \ }0\text{ \ \ \ \ \ \ \ \ \ \ \ \ \ \ \ \ \ \ \ \ \ \
\ \ \ \ \ \ \ \ \ \ \ \ \ \ \ \ \ \ \ \ \ \ }\varepsilon \leq 0
\end{array}
\right. \text{ ,}  \label{eq2}
\end{equation}
where $q$ is a real parameter. This distribution function incorporates King's
distribution function and recovers it in the limit $q\rightarrow 1$. For simplicity,
we also assume that the stars all have the same mass and the velocity
distribution is isotropic everywhere. The dependence on $r$ is determined by
Poisson equation

\begin{equation}
\nabla ^{2}\Psi =-4\pi G\rho \text{ ,}  \label{eq3}
\end{equation}
where $\rho $ is the density given by

\begin{equation}
\rho =4\pi \int_{0}^{\sqrt{2\Psi }}f\left( \Psi -\frac{1}{2}v^{2}\right)
v^{2}\text{d}v\text{ .}  \label{eq4}
\end{equation}
Substituting Eq. (\ref{eq4}) into (\ref{eq3}) we obtain

\begin{equation}
\frac{\text{d}}{\text{d}r}\left( r^{2}\frac{\text{d}\Psi }{\text{dr}}\right)
=-\frac{\left( 4\pi \right) ^{2}G\rho _{1}r^{2}}{\left( 2\pi \sigma
^{2}\right) ^{3/2}}\int_{0}^{\sqrt{2\Psi }}\left\{ \left[ 1+(1-q)
\frac{\varepsilon}{\sigma ^{2}} %
\right] ^{\frac{1}{1-q}}-1\right\} v^{2}\text{d}v\text{ .}  \label{eq5}
\end{equation}
For convenience, we put the last equation in terms of $\Psi /\sigma ^{2}$
and dimensionless quantity $\overline{r}\equiv r/r_{0}$, and it becomes

\begin{equation}
\frac{\text{d}}{\text{d}\overline{r}}\left( \overline{r}^{2}\frac{%
\text{d}}{\text{d}\overline{r}}\left( \frac{\Psi }{\sigma ^{2}}\right)
\right) =-\frac{9\overline{r}^{2}\left\{ \int_{0}^{\frac{\sqrt{\Psi }}{\sigma }}%
\left[ 1+(1-q)\left( \frac{\Psi }{\sigma ^{2}}-u^{2}\right) \right] ^{\frac{1%
}{1-q}}u^{2}\text{d}u-\frac{1}{3}\left( \frac{\Psi }{\sigma ^{2}}\right) ^{%
\frac{3}{2}}\right\} }{\int_{0}^{\frac{\sqrt{\Psi (0)}}{\sigma }}\left[ 1+(1-q)\left( 
\frac{\Psi (0)}{\sigma ^{2}}-u^{2}\right) \right] ^{\frac{1}{1-q}}u^{2}\text{%
d}u-\frac{1}{3}\left( \frac{\Psi (0)}{\sigma ^{2}}\right) ^{\frac{3}{2}}}%
\text{.}  \label{eq6}
\end{equation}
This equation should be integrated numerically. We use the Runge-Kutta
method to calculate it. As usual, we choose the initial condition for $d\Psi
/dr=0$ at $r=0$. Different models can be obtained by giving the central
potential $\Psi (0)/\sigma ^{2}$ and the value of $q$. For $q>1$, the
validity of Eq. (\ref{eq6}) is restricted by the condition $(q-1)<1/\left[
\Psi (0)/\sigma ^{2}\right] $ in order to maintain the quantity inside of
the square brackets positive.

In order to compare the theory with observational data, the density
calculated from Eqs. (\ref{eq4}) and (\ref{eq6}) is substituted into the
projected density $\Sigma (r)$ formula that is given by

\begin{equation}
\frac{\Sigma (r)}{\rho _{0}r_{0}}=2\int_{\overline{r}}^{\infty }\frac{%
\overline{\rho }(\widetilde{r})\text{d}\widetilde{r}}{\left[ 1-\frac{%
\overline{r}^{2}}{\widetilde{r}^{2}}\right] ^{\frac{1}{2}}}\text{ ,}  \label{eq7}
\end{equation}
where $\overline{\rho }(\widetilde{r})=\rho (\widetilde{r})/\rho _{0}$.

We should point out that King models are parameterized by the central
potential $\Psi (0)/\sigma ^{2}$ or the concentration $c=\log _{10}\left(
r_{t}/r_{0}\right) $. In these models, the tidal radius or $c$ is uniquely
determined by the value of the central potential. The greater the value of
the central potential, the greater the tidal radius. In our models, the
concentration is not uniquely determined by the central potential, but we
can adjust the additional parameter $q$. In Fig.1 we show the density
profile for $\Psi (0)/\sigma ^{2}=12$ with different values of $q$. We can
see that the tidal radius increases with the value of $q$. In Fig.2 we show
the surface brightness $I(r)$ of our models, on the scale $\mu =-2.5\log
I(r) $ \ versus $r^{1/4}$, with different values of $q$. We also note that
the curves become shallow with the decrease in the value of $q$, i.e., for $%
q>1$, the curves become more crooked than those of $q=1$, and \ for $q<1$,
the curves become more straight in some range and they become more and more
straight with the decrease in the value of $q$. In the latter case, the
larger the value of $\Psi (0)/\sigma ^{2}$, the longer the straightend part
of the curve becomes. This fact is important to the fitting of some
elliptical galaxies, (for example NGC3379) because the surface brightness of
King models cannot run straight over 10 magnitudes with the increase in
tidal radius and their curves become more crooked (see Fig. 2). In Fig.3, we
fit the surface brightness of galaxy NGC3379. The value of $r_{0}$ is chosen
so that the theoretical curve falls on the observational data. We see that
our theoretical curve can fit the data of the outer region and runs almost
straight over 10 magnitudes, in contrast to the King model which runs less
than 10 magnitudes. In Fig.4 we fit the surface brightness of globular
cluster (47 TUC). We note that the King model (c=2.03) fits tolerably well,
in contrast with our model which fits very well.

In summary, we have extended King models to describe spherically symmetric
systems. Our models are based on the distribution function (\ref{eq2}) and
can be viewed as a truncated Tsallis distribution function. In these models,
we have considered that the stars all have the same mass and the velocity
distribution is isotropic everywhere. We have analyzed the general
properties of these models and contrasted them with King models ($q=1$). As 
applications of our models , we have fitted the surface brightness of
NGC3379 and 47 TUC. From Fig.3, we see that the model with $q\neq 1$ fits
the surface brightness of the galaxy tolerably well, in contrast to the case of $q=1$ 
which cannot fit the outer region. Moreover,
the improvement of our model in relation \ to the King model is clear from
Fig.4. These results suggest that the velocity distribution of stellar systems
described here deviates from Maxwellian shape. Of course, our models
represent the simplest of models, but can be extended to include the
different masses and anisotropy of actual clusters.

\bigskip 

\textbf{Acknowledgment}

\bigskip 

I. T. Pedron is very indebted to CAPES (Brazilian agency) for a scholarship.

\begin{center}
\bigskip \textbf{FIGURE CAPTIONS}
\end{center}

\bigskip

\bigskip

Fig.1 - Density profile plots with $\Psi (0)/\sigma ^{2}=12$ \ and different
values of $q$.

Fig.2 - Surface brightness plots versus $r^{1/4}$ with $\Psi (0)/\sigma
^{2}=8,12$ and different values of $q$. The dotted curves correspond to $%
\Psi (0)/\sigma ^{2}=8$: from bottom to top the value of $q$ is given by $%
q=0.92,0.96,1,1.02,1.04$. While, the full curves correspond to$\Psi
(0)/\sigma ^{2}=12$: from bottom to top the value of $q$ is given by $%
q=0.6,0.8,1,1.02,1.04$.

Fig.3 - Surface brightness fit of galaxy NGC3379 \ measured by different
observers along the galaxy's east-west axis (from data published by
Vaucouleurs and Capaccioli \cite{kormendy}). The dashed curve corresponds to 
$\Psi (0)/\sigma ^{2}=9.2$ and $q=1$, whereas the full curve 
corresponds to $\Psi (0)/\sigma ^{2}=12.2$ and $q=0.939$.

Fig.4 - Surface brightness fit of globular cluster 47 TUC by different
observers (from data published by Illingworth \& Illingworth \cite{kormendy}
and King et. al., 1968 \cite{kormendy}). The dotted curve corresponds to $%
\Psi (0)/\sigma ^{2}=8.868$ and $q=1$, whereas the full curve corresponds to 
$\Psi (0)/\sigma ^{2}=10$ and $q=0.966$.

\end{document}